\documentclass[12pt,preprint]{aastex}

\begin{document}

\title{Absolute Proper Motion of the Canis Major Dwarf Galaxy Candidate}

\author{Dana I. Dinescu\altaffilmark{1,2}, David Mart\'{i}nez-Delgado\altaffilmark{3,4},
Terrence M. Girard\altaffilmark{1},
Jorge Pe\~{n}arrubia\altaffilmark{3}, Hans-Walter Rix\altaffilmark{3},
David Butler\altaffilmark{3}, 
and William F. van Altena\altaffilmark{1}}

\altaffiltext{1}{Astronomy Department, Yale University, P.O. Box 208101,
New Haven, CT 06520-8101 (dana@astro.yale.edu, girard@astro.yale.edu, vanalten@astro.yale.edu)}
\altaffiltext{2}{Astronomical Institute of the Romanian Academy, Str.
Cutitul de Argint 5, RO-75212, Bucharest 28, Romania}
\altaffiltext{3} {Max-Planck Institute fur Astronomie, Heidelberg, Germany 
(jorpega@mpia.de,rix@mpia.de, butler@mpia-hd.mpg.de)}
\altaffiltext{4} {Instituto de Astrofisica de Andalucia (CSIC), Granada, Spain (ddelgado@iaa.es)}

\begin{abstract}
We have measured the absolute proper motion of the candidate 
Canis Major dwarf galaxy (CMa) at $(l,b) = (240\arcdeg, -8\arcdeg)$.
Likely main-sequence stars in CMa 
have been selected from a region in the color-magnitude 
diagram that has very little contamination from known Milky Way components.
We obtain $\mu_{l}~cos~b = -1.47 \pm 0.37$ and
$\mu_{b} = -1.07 \pm 0.38$ mas yr$^{-1}$, on the ICRS system via
Hipparcos stars. 
Together with the radial velocity of 109 km s$^{-1}$,
and the assumed distance of 8 kpc, these results imply a space motion
of $(\Pi, \Theta, W) = (-5 \pm 12, 188 \pm 10, -49 \pm 15)$ km s$^{-1}$.
While CMa has in-plane rotation similar to the mean of thick disk stars, 
it shows significant $(3\sigma)$ motion perpendicular to the disk, and
differs even more $(7\sigma)$ from that expected for the Galactic warp.
The $W$ velocity lends support to the argument that the CMa overdensity
is part of a satellite galaxy remnant.
\end{abstract}


\section{Introduction}

The Sloan Digitized Sky Survey team discovered
a low-metallicity ([Fe/H] $\sim -1.6$), low-Galactic-latitude structure 
toward the constellation of Monoceros that extends over 50 degrees in the sky,
at $\sim 20$ kpc-distance from the Galactic center
(Newberg et al. 2002, Yanny et al. 2003). Subsequent observations 
by Ibata et al. (2003) show that this structure extends over 100 degrees,
in a ring-like configuration with distances between 
15 and 20 kpc from the Galactic center. Using 2MASS M-giant tracers,
Rocha-Pinto et al. (2003) confirmed the previous findings,
 and indicated the presence of a 
metal rich population ([Fe/H] $\sim -0.5$) as well.
This ``Monoceros (Mon) stream''
is believed to be the tidal stream of a disrupting and merging dwarf galaxy.
Evidence in favor of this picture are the
spatial distribution and the velocity dispersion of stars in the stream
(Crane et al. 2003).

%A number of globular clusters
%were also hypothesized to belong to the Mon stream (Frinchaboy et. al. 2003,
%Martin et al. 2004a). For those clusters that have proper motions, their 
%orbits are not consistent with dynamical membership to the 
%Mon stream (Pe\~{n}arrubia et al. 2005, P05).

Martin et al. (2004a) discovered a seemingly distinct stellar 
overdensity in the constellation of Canis Major, from the mapping of the sky 
with  2MASS-selected M giants.
Deep color-magnitude diagrams in this region (Bellazzini et al. 2004 - B04,
and Mart\'{i}nez-Delgado et al. 2005 - M05) show a well-defined
main sequence, as well as a plume of blue stars which are believed to be
main sequence stars younger (1-2 Gyr) than the main population of the 
overdensity (4-10 Gyr). Martin et al. 2004a suggest that the 
Canis Major overdensity
is the core of the progenitor of the Mon stream. Located at 
$(l,b) = (240\arcdeg, -8\arcdeg)$, this overdensity has also been interpreted as due
to the Milky Way warp and flare by Momany et al. (2004). These authors
used UCAC2 (Zacharias et al. 2004) proper motions to argue
that the motion of the CMa overdensity is indistinguishable 
from that of the thin disk. 
However, the uncertainty in the UCAC2-derived mean motion 
($\sim 2$ mas yr$^{-1}$) is a weak discriminant among these hypotheses.
Pe\~{n}arrubia et al. (2005, P05) have performed N-body simulations 
constrained by the available spatial and kinematical information
for the Mon stream to test whether CMa is the 
progenitor of the Mon stream. They could only determine that CMa is 
a plausible, perhaps even likely progenitor of the Mon stream.
While the narrowness of the main sequence of the CMa overdensity
(B04 and M05) and the 
radial velocities in CMa (mean and dispersion, Martin et al. 2004b) argue
in favor of it being a dwarf remnant rather than the thin disk warp, the common origin 
of the Mon stream and CMa can be tested stringently with a
precise measurement of its absolute proper motion\footnote[1]{For simplicity, we will refer to the Canis Major
overdensity as the Canis Major dwarf, although
we are aware that
there isn't yet consensus regarding the nature of this
stellar overdensity (i.e., core of the satellite dwarf, stream of the satellite
or Galactic warp).}. 
It is the purpose of the present work to provide such a proper-motion
measurement.
%The proper-motion determinations are described in Sections 2 and 3.
%Section 4 includes a brief discussion of the result.

\section{Proper-Motion Determinations}

Our proper-motion measurement is based on two photographic plate pairs that
are part of the Southern Proper Motion Program (SPM, e.g., Girard et al. 2004)
and on CCD observations made with the Wide-Field Imager (WFI) at prime focus of 
the 2.2m ESO/MPG telescope 
at La Silla Observatory in December 2003 (M05). 
The SPM plates were taken
with the 50-cm double astrograph at Cesco Observatory in El Leoncito,
Argentina (plate scale = 55.1 $\arcsec$/mm), in 1967 and in 1988.
At each epoch, two plates are taken in the $V$ and $B$ passbands, and the
area covered is $6.3\arcdeg$x$6.3\arcdeg$. The CCD data cover an 
area of about half a degree on a side, and they are located on a corner of the 
SPM field, at $(l,b) = (240.15\arcdeg, -8.08\arcdeg)$. 
Each SPM plate contains a two-hour exposure, which
reaches to about $V = 18$, and an offset two-minute exposure,
which allows a tie-in to bright reference stars. During both exposures, an
objective grating is used, which produces a series of diffraction images
on either side of the central zero-order image. 
The presence of the grating images is advantageous for two reasons:
It effectively extends the dynamic range of the plates, in both the long
and short exposures, providing unsaturated images of brighter stars.
More importantly, the grating images provide the means for calibrating and 
correcting the magnitude equation in each exposure.
(See Girard et al. 1998 for a detailed discussion of the techniques involved.
The theory is described in Section 3 of that paper, and Section 4
describes its application to SPM plates as well as an assessment
of its effectiveness.)

The photographic plates were
scanned  with the Yale PDS microdensitometer, at a pixel size of 12.7 microns.
The preparation of the 
input list of stars to be measured, the PDS measurements, and the astrometric
reductions are all similar to those described by Dinescu et al. (2003), 
Girard et al. (1998) and 
references therein. 
%Here, we will only briefly outline the procedure. 
%We have measured a selected set of stars over the whole area of the
%SPM field. This set  includes Hipparcos and Tycho stars (ESA 1997) that provide
%the correction to absolute proper motion and the photometric calibration
%respectively. Tycho stars together with an additional 
%subset of $\sim 2100$ Guide Star Catalog 1.1 stars 
%(GSC, Lasker  et al. 1990) assure a good range in the magnitude
%of various diffraction orders in order to model magnitude-dependent
%systematics (Girard et al. 1998). 
%For all of these relatively bright stars, we measure
%long and short-exposure images, and diffraction images. 
%A total of 2800 faint, anonymous stars selected from the USNO-A2.0 catalog
%(Monet  et al. 1998) were also measured in order to model the plate 
%transformations. 
In the area of the plate covered by the 
CCD observations we measure all of the stars listed in the CCD data 
that are detectable
on the photographic plates.
%The CCD data (the short exposure listed in
%M05, their Fig. 1) 
%reach a much fainter limit than our $V\sim 18$ plate limit.

We have performed two types of proper-motion solutions that
eventually are combined to produce the final result. The first one
is a separate reduction of the visual and blue plate pair, and 
excludes the CCD observations. This determination, in principle, can provide 
two independent measurements 
of the {\it absolute} proper motion of CMa. However, because  CMa candidate
stars are faint ($V \ge 16$), their individual proper motions will have large
proper-motion uncertainties (see Girard et al. 2004). As we expect
only $\sim 150$ CMa candidates within the CCD region,
 their mean motion will be poorly
determined. Consequently, we will use this plate-pair approach 
only to determine the correction to absolute proper motion via Hipparcos stars,
and to tie into a new proper-motion system (see below).
The second approach combines all four plates and the CCD positions
to determine {\it relative} proper motions for the stars in the area
limited by the CCD data. In this way the time baseline is extended
from 21 years to 36 years with the immediate result of improving
the precision of the proper motions.
The ``all-plate'' proper-motion system is tied to each ``plate-pair''
proper-motion system using proper-motion offsets. These offsets
are the means of the proper-motion differences 
between the two systems for about 1600 stars for the visual pair and
1700 stars for the blue plate pair.

%Each photographic plate is pre-corrected for differential refraction,
%and magnitude-dependent systematics as modeled by diffraction images
%(Girard et al. 1998). 
For the plate-pair solution, different-epoch plates
are mapped into each other using the faint anonymous stars, 
and a polynomial plate model that includes only geometric terms.  
(Magnitude equation was corrected previous to the plate-pair solution
using the techniques referenced earlier.)
The correction to absolute proper motion is defined by the motion of
$\sim 120$ Hipparcos stars with respect to the faint anonymous (reference) stars.
First, the mean proper-motion of Hipparcos stars is determined for
each image order (there are a total of 2 central orders for 
the long and short exposures
and 4 diffraction orders; symmetric positions of the same order are averaged). 
The final value for each plate pair 
is a weighted mean of all image-order values. The weights are given by the
internal scatter within each image order.
For the blue plate pair we obtained $\mu_{\alpha}~cos~\delta = -1.30 \pm 0.24$
and $\mu_{\delta} = 3.34 \pm 0.36$ mas yr$^{-1}$. For the visual plate pair 
we obtained $\mu_{\alpha}~cos~\delta = -1.79 \pm 0.19$
and $\mu_{\delta} = 0.90 \pm 0.19$ mas yr$^{-1}$. 
We note that the blue and visual values need not be in agreement since,
on average, different faint reference stars were used in each plate-pair 
solution.
 
The all-plate solution combined all positions in one proper-motion solution,
in an iterative process (e.g., Girard et al. 1989). 
The CCD data are from 8 individual chips of the WFI, which 
were independently reduced into the photographic plates. 
Formal individual uncertainties increase with magnitude
from 1.4 mas yr$^{-1}$ at $V \sim 12$-15 up to 4  mas yr$^{-1}$ 
at $V \sim 17.5$. The proper-motion offsets between the system of 
the all-plate solution and that of the
blue plate pair are: $\mu_{\alpha}~cos~\delta = -0.12 \pm 0.12$
and $\mu_{\delta} = -1.96 \pm 0.11$ mas yr$^{-1}$. Similarly, 
for the visual plate pair we obtain:  $\mu_{\alpha}~cos~\delta = 0.32 \pm 0.13$
and $\mu_{\delta} = 0.47 \pm 0.13$ mas yr$^{-1}$. The final correction 
to be applied to the relative proper motions of the all-plate system 
in order to transform them to absolute proper motions is the sum 
of the system offsets and the proper motions of Hipparcos stars with
respect to a given plate-pair reference system. These summed corrections are:
for the blue plate pair,  $\mu_{\alpha}~cos~\delta = -1.42 \pm 0.27$, 
$\mu_{\delta} = 1.38 \pm 0.38$ mas yr$^{-1}$, and for the visual plate pair,
$\mu_{\alpha}~cos~\delta = -1.47 \pm 0.23$, $\mu_{\delta} = 1.37 \pm 0.23$ 
mas yr$^{-1}$. We take the 
error-weighted average of the two determinations as the final value
of the correction to absolute proper motion of the all-plate system. 
This is $\mu_{\alpha}~cos~\delta = -1.45 \pm 0.18$, 
$\mu_{\delta} = 1.37 \pm 0.20$ mas yr$^{-1}$.
Here we assume that the two determinations
are independent, although strictly speaking 
they are not, because the system proper-motion offsets are highly correlated.
However, the dominant source 
of uncertainty comes from the Hipparcos tie-in measurement which is
determined independently from two plate-pairs.

\section{Absolute Proper Motion of CMa}

In Figure 1 (top panel) we show the color-magnitude diagram (CMD) for all
stars from M05 that were successfully 
measured on the photographic plates. The photometry is that from 
M05. 
The bottom panel of Fig. 1 shows a simulated CMD
from the Besan\c{c}on Galactic model (Robin et al. 2003) in the
 area of the sky covered by the CCD 
observations. This model includes the stellar warp as characterized from
preliminary results from the DENIS survey (Epchtein et al. 1997).
Two regions of interest are marked in each panel:
the ``blue plume'' (BP) region likely representing the upper main sequence 
of a younger sub-population in CMa, with very little Milky-Way contamination,
and the ``red-clump'' (RC) region of CMa ($1.4 \le B-R \le 1.8$, and $14.5 \le R \le 16$).
These regions are described in detail below.

There is a $B-R$ color offset of $0.19$ between the observations 
and the model in the sense that the
simulated data are redder than the observations. This may be attributed to the 
particular reddening used in the Besan\c{c}on Galactic model.
Consequently, we have shifted the simulated data by this offset. 
The model data are used only for a qualitative understanding 
of the expected Galactic populations and not a quantitative comparison.
The overall shapes of the CMDs are similar. Toward the red end, the 
incompleteness of the observations is apparent 
when compared to the Galactic model. 
Bright stars ($R < 14.5$) are also missing from Fig. 1, top panel, 
as they were saturated in the initial input 
list derived from CCD data. The main difference between the observations 
and the Galactic model
are the blue, faint stars ($16 < R < 18$, and $ B-R \le 0.8$) already noted
by B04 and M05. 
%These stars are 
%suspected to be a main-sequence population younger than the majority of stars
%in the CMa dwarf. 
As in the above-mentioned papers, we will refer to it as the blue plume (BP).
For magnitudes brighter than $R\sim 16$,
 there is a significant number
of stars in the Galactic model, blueward of the halo/thick disk main 
sequence turn-off. 
According to the model, these are 2-3 Gyr-old main sequence stars at distances
$\sim 2-4$ kpc. Their model velocities indicate that they belong to a
thin-disk like population.  

To avoid Galactic star contamination, we have drawn
a polygon (highlighted in Fig. 1, top panel) guided by the configuration of the
Galactic-model CMD. Only stars within this polygon were considered in the proper-motion
determination, and these are the candidate BP stars of CMa.
Previous attempts to estimate the motion of CMa
from radial velocities, proper motions and the spatial configuration (e.g.,
Momany et al. 2004, Martin et al. 2004b, P05 and references there in)
 indicate that its orbit is  prograde and
not highly inclined with respect to the disk.
If this is the case, the proper motion of 
CMa will overlap with the proper-motion distribution of distant 
disk stars that will have 
small proper motions and a low proper-motion dispersion. 
%For example,
%a disk velocity dispersion of $\sim 20$ km s$^{-1}$ translates into 
%a proper-motion dispersion
%of 2 mas yr$^{-1}$, at a distance of 2 kpc. This proper-motion dispersion
%is comparable to our proper-motion uncertainties per star. 
%Therefore, it is not 
%possible to separate CMa stars from disk stars in proper-motion space alone.
Samples that include both CMa stars and
Galactic disk stars, such as the 
red clump region (RC) of CMa (see Fig. 1) and the giant branch of CMa 
seen in the 2MASS studies, will be subject to proper-motion bias from 
the disk stars. 
To illustrate, in Figure 2 we show 
the absolute proper-motion distribution of the observations 
(top) and of the Galactic model predictions (bottom) in
the CMa red-clump region of the CMD.
The Galactic-model proper motions were convolved with 
proper-motion uncertainties that mimicked those of our observations.
The two distributions are similar, and it is difficult
to decide if a kinematically distinct population is present
 in the observations that is not in the Galactic model.
In the model, the stars that appear to clump
(i.e., have a proper-motion dispersion comparable 
to the proper-motion uncertainties) are primarily $\sim 5$ Gyr-old 
disk giants at distances between 
3 and 8 kpc. It is this population that will bias the CMa
proper motion in samples that use this portion of the CMD.
Therefore it is crucial to select photometrically a clean sample of CMa 
candidates to determine the proper motion.

In Figure 3 we show the absolute proper motions of CMa candidates selected
from the BP polygon region in Fig. 1.
The single-coordinate
proper-motion error estimates vary from 2.2 to 4.1 mas yr$^{-1}$ as
a function of magnitude, with a mean value of 3.1 mas yr$^{-1}$.
Also shown are the proper-motion marginal distributions 
along the two axes.
We have chosen to use a trimmed mean to estimate the average motion
because of the apparent presence of outliers.
Probability plots (Lutz \& Upgren 1980) indicate that the inner 80\%
of the sample is roughly Gaussian.
This corresponds to 104 (of the total 130) stars within 8 mas yr$^{-1}$ of
the tentative mean, limits that also appear reasonable based on
the appearance of the marginal distributions.
From these 104 stars, the mean absolute proper motion of CMa is 
$\mu_{\alpha}~cos~\delta = -1.61 \pm 0.38$, 
$\mu_{\delta} = 0.84 \pm 0.37$ mas yr$^{-1}$. 
In Galactic coordinates, it is $\mu_{l}~cos~b = -1.47 \pm 0.37$, 
$\mu_{b} = -1.07 \pm 0.38$ mas yr$^{-1}$.
The quoted uncertainties include the contribution from
the correction to absolute proper motion. 
The dominant source of uncertainty is the determination of the mean
for these faint BP stars. 
The estimated mean is indicated in Figure 3 by the intersecting solid lines.

\section{Discussion}

Taking the distance to the CMa core to be $8.1 \pm 0.4$ kpc 
(M05, B04), and the radial velocity to be 
$109 \pm 4$ km s$^{-1}$
(Martin et al. 2004b), we obtain the velocity with respect to the 
Galactic rest frame in cylindrical coordinates:
 $(\Pi, \Theta, W) = (-5 \pm 12, 188 \pm 10, -49 \pm 15)$ km s$^{-1}$.
Assumptions include the peculiar Solar motion from Dehnen \& Binney (1998),
 that the Sun is located at 8.0 kpc from the
Galactic center (GC), and the rotation of the local standard of 
rest is 220.0 km s$^{-1}$.
Martin et al. (2004b) determine two peaks in the radial velocity
 distribution of
M giants in CMa. Both peaks have a low radial velocity dispersion 
(i.e., 9 and 13 km s$^{-1}$), indicative of a common-motion, cold system. 
We adopt the mean of the most
populated peak as the radial velocity of the CMa dwarf. 
P05 models
also favor this value, while the second peak may be due to a 
tidal tail surrounding CMa.
In the Galactic rest frame, the proper motion is:
$\mu_{l}~cos~b' = -4.16\pm 0.37$, $\mu_{b}' = -1.61 \pm 0.38$ mas yr$^{-1}$.
%Given the location of the CMa dwarf, the $W$ velocity component is
%almost entirely determined by the proper motion along Galactic latitude
%(i.e., $W$ scales with distance). 
 
Momany et al. (2004) suggest that the warping of the disk is
responsible for the CMa overdensity.
The geometry of the stellar warp and flare have been recently parameterized
by L\'{o}pez-Corredoira et al. (2002). The line of nodes 
of the warp is $\sim 5\arcdeg$ away from the GC-Sun direction, in the sense
of Galactic antirotation. The maximum 
deflection below the Galactic plane is in quadrant 4, and  
above the plane in quadrant 1 (see Fig. 22 in L\'{o}pez-Corredoira 
et al. 2002). 
%At maximum amplitude, stars in the warp have 
%$W = 0$ km s$^{-1}$, while, along the line of nodes,
%$W$ will be maximum (positive and negative).
In quadrant 3, where CMa is located, material from the warp 
is moving from below the plane, up toward the plane, i.e., with 
a positive $W$ velocity. 
%It is only on the far side of the bulge, 
%that the warp has a negative $W$ velocity component. 
Following Drimmel, Smart \& Lattanzi (2000, D00), where the kinematic 
signature of the warp is derived for a traditional, long-lived Galactic
warp, (their equations 5 and 11, and no precession),  
we obtain $W_{warp} = 52$ km s$^{-1}$. 
%Interestingly, D00 measure
%a negative $W$ velocity component for a sample of
%distant OB stars, in disagreement with the predicted motion of the long-lived
%warp. 
In order to obtain a negative $W$ velocity, D00 had to invoke 
an unphysically high precession rate for the warp.
Our $W$ velocity component is inconsistent ($7\sigma$)
with the expected motion of the warp at this Galactic location.
If there is no warp, then the $W$ velocity of CMa is $3.3\sigma$ 
different from the velocity of the thin disk.
The motion of the warp as estimated above, $(\Pi, \Theta, W) = (0, 220, 52)$ km s$^{-1}$, translates into 
$\mu_{l}~cos~b' = -4.95$, $\mu_{b}' = 0.95$ mas yr$^{-1}$, at the CMa distance.
%It is thus the proper-motion component along Galactic latitude that is crucial
%in distinguishing between the disk/warp motion and that of a system
%of an external origin.
The hypothetical warp motion, transformed to celestial coordinates, is
indicated in Figure 3 as the intersecting dotted lines.

By integrating the orbit in the Johnston, Spergel \& Hernquist (1995)
 Galactic potential model, we obtain 
a pericenter of $10.5 \pm 0.9$ kpc, and an apocenter of $14.0 \pm 0.2$ kpc. 
The maximum distance from the Galactic plane is 
$2.0 \pm 0.4$ kpc, and the orbital period is $342 \pm 14$ Myr.
The orbit inclination is $ 15\arcdeg \pm 3\arcdeg$ and the eccentricity is $0.14 \pm 0.04$. 
The uncertainties in the orbital elements were derived 
in a Monte Carlo fashion from the uncertainties in the distance, 
radial velocity and proper motions (Dinescu, Girard \& van Altena 1999).
The derived  motion makes CMa a likely
parent of the Mon stream as envisioned by P05.
Indeed, their modeling gives an orbital inclination of 
$20\arcdeg \pm 5\arcdeg$, and
an eccentricity of $0.10 \pm 0.05$ for the Mon stream progenitor.
Currently, CMa is at apocenter, and
N-body simulations show that it should undergo tidal disruption (P05).
The original orbit of the progenitor seems to have been highly coupled
with the disk motion, i.e., a prograde, low-inclination orbit of 
low eccentricity. Therefore the interaction with the disk could have 
altered the orbit of the progenitor for instance, by decreasing 
the initial orbital inclination (Walker, Mihos \& Hernquist 1996).
Likewise, this interaction may have altered the shape of the disk,
perhaps inducing the existent warp.

We acknowledge funding from the National Science Foundation, grant AST-0407293.

\clearpage

\begin{figure}
\plotone{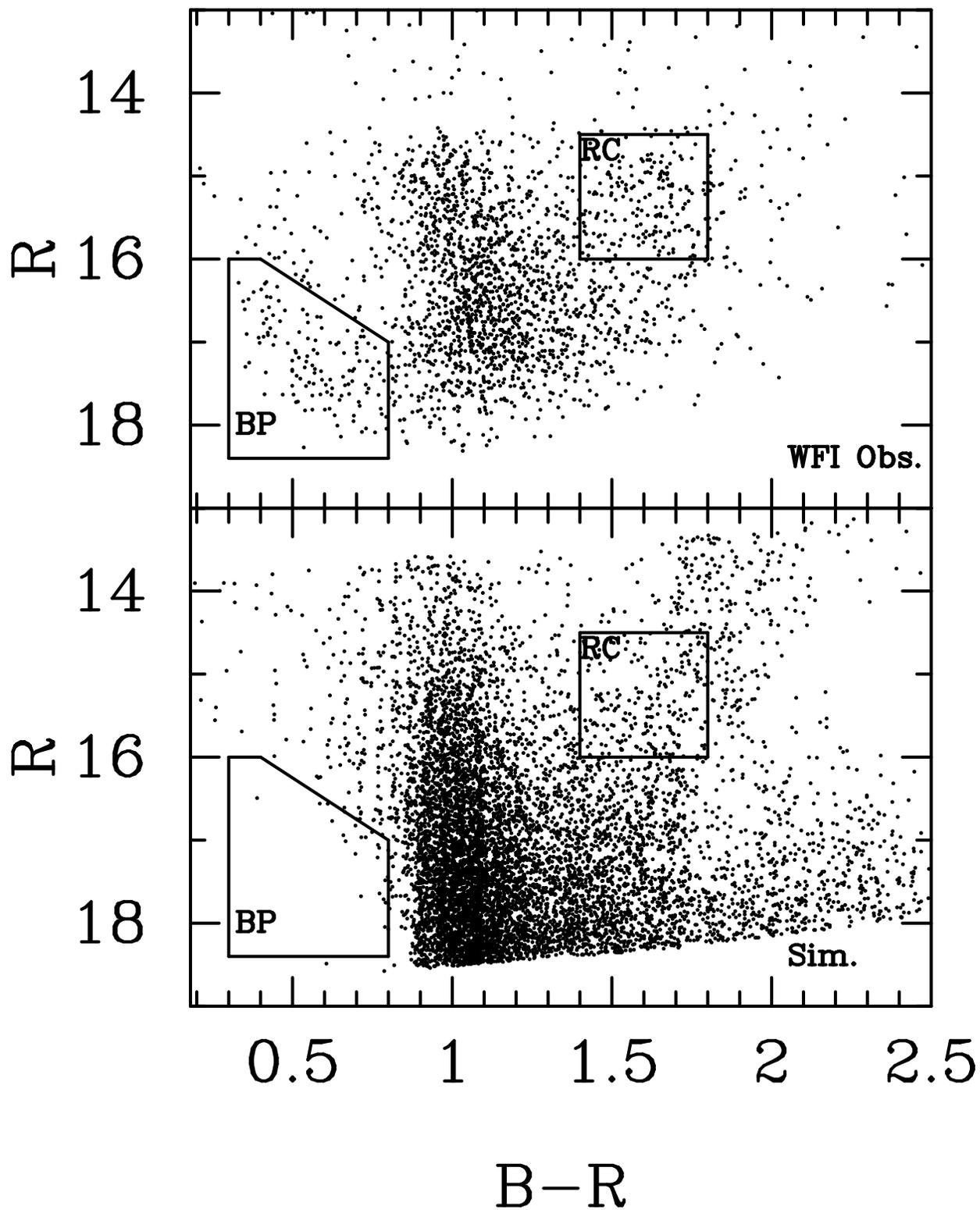}
\figcaption{Color-magnitude diagrams of the stars observed in the CMa region 
with proper-motion measurements (top), and the 
simulations of known Milky-Way components(bottom). 
The blue-plume region and the red-clump region 
are marked. A 0.19 $B-R$ color shift has been applied to the 
simulated data (see text).}
\end{figure}

\clearpage

\begin{figure}
\epsscale{.60}
\plotone{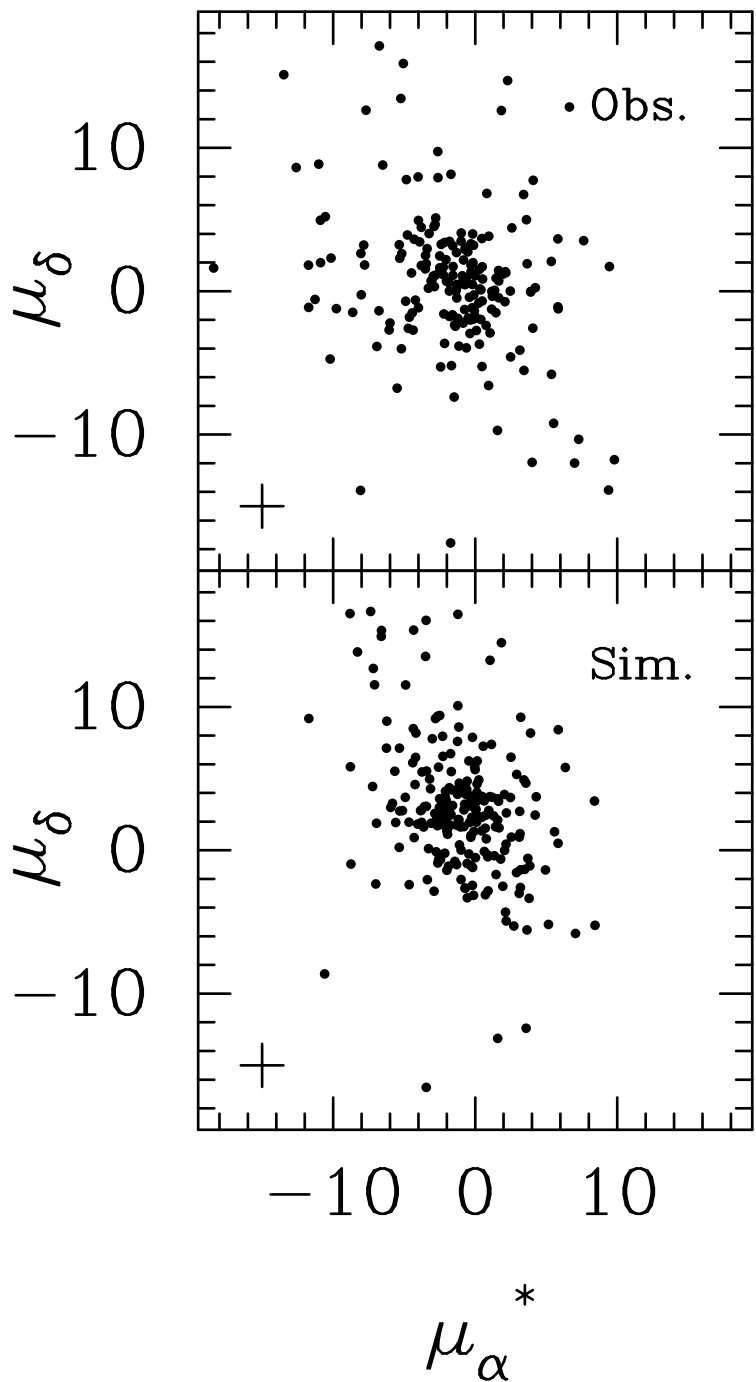}
\figcaption{Vector-point diagram of stars in the red-clump region 
of CMa. Observations are shown in the top panel, simulated data 
for the smooth Milky Way components in the same direction are shown
in the bottom panel ($\mu_{\alpha}^* = \mu_{\alpha}~cos~\delta$). 
Typical proper-motion uncertainties are indicated in each panel. It is apparent
that CMa stars can not be separated from disk stars, on the basis of proper motions alone.}
\end{figure}

\clearpage

\begin{figure}
\epsscale{1.0}
\plotone{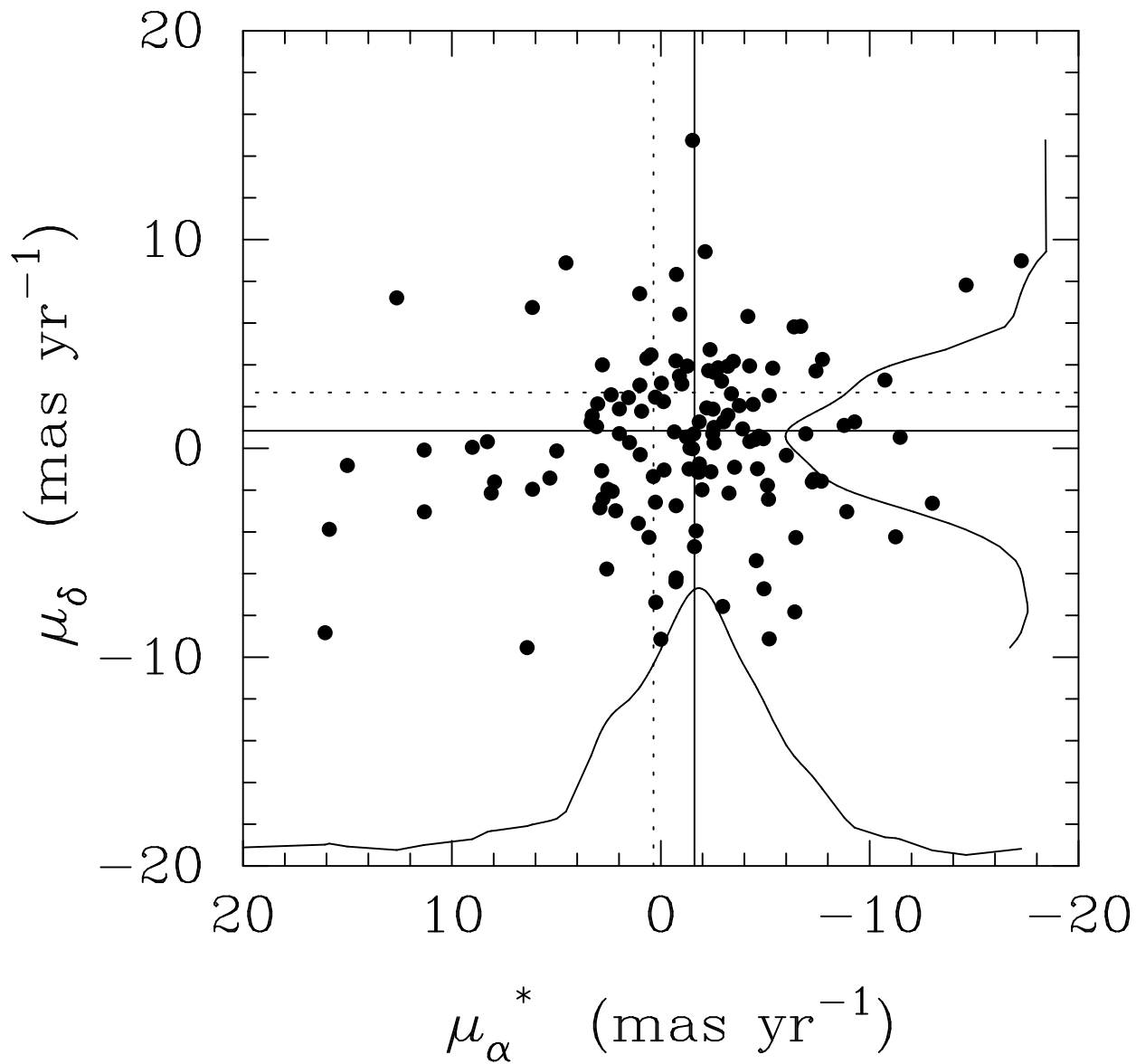}
\figcaption{Absolute proper-motion vector-point diagram of the CMa 
blue-plume candidates.
One-dimensional marginal distributions, also shown, were constructed by
convolving with a 1 mas yr$^{-1}$ Gaussian kernel.
The intersecting solid lines indicate the estimated mean motion of
the sample.
The dotted lines indicate the expected motion of the Galactic warp at
this position.
}
\end{figure}

\end{document}